\documentclass[conference]{IEEEtran}
\IEEEoverridecommandlockouts
\usepackage{cite}
\usepackage{amsmath,amssymb,amsfonts}
\usepackage{textcomp}
\usepackage{xcolor}
\usepackage{cite}
\usepackage{graphicx,color}
\usepackage{textcomp}
\usepackage[hidelinks]{hyperref}
\usepackage{algorithm,algorithmic}
\def\BibTeX{{\rm B\kern-.05em{\sc i\kern-.025em b}\kern-.08em
    T\kern-.1667em\lower.7ex\hbox{E}\kern-.125emX}}
\AtBeginDocument{\definecolor{tmlcncolor}{cmyk}{0.93,0.59,0.15,0.02}\definecolor{NavyBlue}{RGB}{0,86,125}}
\usepackage{array}
\usepackage{subcaption}
\usepackage[hyphenbreaks]{breakurl}
\usepackage{balance}
\def\BibTeX{{\rm B\kern-.05em{\sc i\kern-.025em b}\kern-.08em
    T\kern-.1667em\lower.7ex\hbox{E}\kern-.125emX}}
\begin{document}

\title{GATSY: Graph Attention Network for \\Music Artist Similarity}

\author{%
  \IEEEauthorblockN{%
    Andrea Giuseppe Di Francesco\textsuperscript{\textsection}\IEEEauthorrefmark{1}\IEEEauthorrefmark{2},
    Giuliano Giampietro\IEEEauthorrefmark{1},
    Indro Spinelli\IEEEauthorrefmark{3}, 
    Danilo Comminiello\IEEEauthorrefmark{5}%
  }%
}

\maketitle

\begingroup
  \renewcommand\thefootnote{\fnsymbol{footnote}}
  \footnotetext[1]{Department of Computer Science, Control and Management Engineering, Sapienza University of Rome, Rome, Italy.}
  \footnotetext[2]{Institute of Information Science and Technologies "Alessandro Faedo" - ISTI-CNR, Pisa, Italy.}
  \footnotetext[3]{Department of Computer Science, Sapienza University of Rome, Rome, Italy.}
  \footnotetext[5]{Department of Information Engineering, Electronics and Telecommunications, Sapienza University of Rome, Rome, Italy.}
  \renewcommand\thefootnote{\textsection}
  \footnotetext{Corresponding author, \texttt{difrancesco@diag.uniroma1.it}.}
  \renewcommand\thefootnote{\arabic{footnote}} 
\endgroup

\maketitle
\begin{abstract}
The artist similarity quest has become a crucial subject in social and scientific contexts, driven by the desire to enhance music discovery according to user preferences. Modern research solutions facilitate music discovery according to user tastes. However, defining similarity among artists remains challenging due to its inherently subjective nature, which can impact recommendation accuracy. This paper introduces GATSY, a novel recommendation system built upon graph attention networks and driven by a clusterized embedding of artists. The proposed framework leverages the graph topology of the input data to achieve outstanding performance results without relying heavily on hand-crafted features. This flexibility allows us to include fictitious artists within a music dataset, facilitating connections between previously unlinked artists and enabling diverse recommendations from various and heterogeneous sources. Experimental results prove the effectiveness of the proposed method with respect to state-of-the-art solutions while maintaining flexibility. The code to reproduce these experiments is available at \url{https://github.com/difra100/GATSY-Music_Artist_Similarity}.
\end{abstract}
%


\maketitle

%
%
%
%
%
\section{INTRODUCTION}
\label{sec:intro}
\IEEEPARstart{T}{he} artist similarity task involves finding a match between artists based on their shared aspects. However, there is no unambiguous way to define how two or more artists can be similar and the main characteristics that make them similar \cite{artistQUEST}. Similarity can be defined according to several points of view, such as cultural perspectives (e.g., provenance, historical context) or aspects more commonly related to the content of songs (e.g., common musical progressions, song keys). The ground truth for artist similarity does not rely solely on these individual features but rather on a combination of them.
%
Such an approach is often implemented by extracting and elaborating music tracks. However, the resulting features cannot represent a subjective opinion concerning an artist's categorization. Furthermore, \textit{digitalizing} user tastes is always a critical task that inevitably affects a music recommendation system.
In recent years, the research community has developed various artist similarity methods using Graph Neural Networks (GNNs) \cite{GNNbook}, owing to their natural ability to extract patterns from graph-structured data. Modeling artist relationships through graphs allows not only for the derivation of similarity scores but also for addressing challenges associated with long-tail artist recommendations \cite{artistS2, LongTailMIR, heterogeneousGNNas}. However, generating the necessary artist data to train a GNN often requires information that is not readily available and may involve significant effort to collect, as highlighted in \cite{artistS}. Furthermore, beyond the methodologies designed for data extraction, there are limited open-source datasets available for research purposes.
In this work, we build on the idea of using GNNs for artist similarity and introduce a recommendation system that addresses these challenges. Our approach defines similarity via content-based and cultural-based aspects \cite{artistS}. Experimental results show comparable accuracy to existing models while significantly reducing the number of parameters and offering a more efficient solution. The main contributions of this paper are as follows:
\begin{itemize} 
    \item We introduce \textbf{GATSY} (Graph ATtention network for music artist SimilaritY), a novel graph attention network-based model for artist similarity.
    
    \item We leverage both graph structure and input features, demonstrating robust performance even in real-world scenarios where input node features are missing or incomplete, and connections between nodes are sparse or disconnected.

    \item We augment the \textbf{Olga} dataset \cite{artistS} incorporating artist labels from diverse sources.
    \item We incorporate music genre information into the artist representation, enabling genre-aware similarity while maintaining high accuracy.
    \item We present empirical evidence that adding genre information does not significantly enhance the model ability to estimate artist similarity, suggesting that similarity is not primarily genre-dependent.

    \item We demonstrate how GATSY can serve as the backbone for an artist recommendation system, showcasing its applicability in real-world scenarios.
\end{itemize}


\section{Related Works}
\label{sec:relatedworks}
\textbf{Artist Similarity. }Several methods recently attempted to address the artist similarity problem despite the difficulty in expressing an objective meaning of this similarity \cite{artistQUEST}. Natural language processing (NLP) methodologies found a role of relevance in this scenario by embedding a vast amount of text data describing artists and their songs. For instance, in \cite{semanticSimilarity}, text features were used to define a knowledge base of artists. Also, tasks like the Named Entity Recognition (NER) \cite{NER} were applied to Wikipedia sources to obtain vector representations \cite{SSARGNN}. Besides relying on text attributes, artists have also been represented through their content-based features as in \cite{multidim, heterogeneous}. In particular, in \cite{multidim} it is also shown how heterogeneous features can outperform models with feature-specific approaches. \\
\textbf{Graph Neural Networks. }
In the past decade, GNNs have contributed heavily to several areas of Deep Learning (DL), such as Recommendation Systems \cite{recsystem1,recsystem2}, Natural Language Processing \cite{NMT1, NMT2, q&a1, q&a2, textclass,textclass2}, Computer Vision \cite{VQA, imclassification, imclassification2, objdetect}, and Knowledge graphs \cite{KBcompletion, KGalignment}. Generally, most of the GNNs follow the message-passing formalism \cite{messagepassing}, where nodes update iteratively their representations, after being trained on downstream tasks such as  \textit{link prediction}, and \textit{node} or \textit{graph classification} \cite{GNNbook}. Due to the natural representation of music artists as a network, GNNs have also found a place in artist similarity applications. \\
\textbf{Artist Similarity with Graph Neural Networks. }\cite{artistS} is the seminal work that merges together artist similarity and GNNs. Here, a graph of artists is employed, where user tastes specify the connections, and each artist has an individual representation that is derived from track-level statistics about the loudness, dynamics, and spectral shape of the signal. This work shows that through an unsupervised training strategy, the artist relationships and the individual features help relate effectively to non-previously connected artists. In an extension of this work, namely \cite{artistS2}, it has been proved that their approach also allows for long-tail artist recommendations, where some nodes present few connections w.r.t. those associated with popular artists. However, their approach heavily relies on the knowledge of input features for the artists, which are not always known in advance. Moreover, do not describe the effects of using more expressive GNNs, limiting their analysis to GraphSAGE \cite{inductive} as a GNN candidate. 
GNNs have been employed to solve long-tail artist recommendations also in \cite{LongTailMIR}. More recently, in \cite{heterogeneousGNNas}, multimodal representations and heterogenous networks are used to capture more intricate artist relationships, considering explicitly also \textit{Lyrics}, and \textit{Audio}. This approach formalizes the artist similarity computation as a Link Prediction task, where we estimate the probability that an edge exists between two nodes that resemble two artists.
In this work, we also address artist similarity with GNNs and follow an approach similar to \cite{artistS}, since we compare on the same experimental setup. We show that choosing a more expressive and efficient (in terms of parameters) architecture is a way to overcome the need for hand-crafted artist features.

%
%
%
%
%
%
%
%
\section{The Proposed GATSY Framework}
\label{sec:problem}

\subsection{Addressing Artist Similarity with Graph Neural Networks}

Let $\mathcal{G} = \{\mathcal{V}, \mathcal{E}\}$ be a graph, where $\mathcal{V}$ is the set of nodes, and $\mathcal{E}$ is the set of edges. Being $|\mathcal{V}| = n$ the number of nodes, we define $\mathcal{X} \in \mathcal{R}^{n \times m}$ as the instance matrix containing the artist features, and let $\mathcal{A} \in \{0,1\}^{n \times n}$ describe the symmetric adjacency matrix of the graph. Following the approach proposed in \cite{artistS}, $\mathcal{X}$ brings content information about the music of artists while $\mathcal{A}$ provides the relational information. Given these matrices, GNNs can produce an embedding of artists. In this sense, a GNN works as an encoder, which can be designed in order to derive the similarity between samples. Denoting GNN$(\cdot,\cdot)$ as a stack of generic \textit{graph convolutional} (GC) layers interleaved by nonlinearities, we can define $\mathcal{Z} \in \mathcal{R}^{n \times m'}$ to represent the model output, where $m'$ is the embedding dimension:
\begin{equation}
\label{eq1}
    \mathcal{Z} = \text{GNN}(\mathcal{X}, \mathcal{A}).
\end{equation}

\noindent Each vector $\mathbf{z}_i \in \mathcal{Z}$ exhibits the latent representation of the $i$-th artist.\\
Let us also assume the knowledge of $\mathcal{Y} \in \{0,1\}^{n \times c}$, which encompasses the ground truth labels $\mathcal{C}$, s.t. $|\mathcal{C}| = c$, associated with each sample. Starting from $\mathcal{Z}$ we can predict the artist labels with a \textit{multi-layer perceptron} (MLP) as follows. 
\begin{equation}
    \label{eq:supervised}
    \hat{\mathcal{Y}} = \text{MLP}(\mathcal{Z}).
\end{equation}

In \cite{artistS}, GraphSAGE layers \cite{inductive} have already been proposed as GNN$(\cdot, \cdot)$. In this work, we investigate other GNNs \cite{conf1, GIN}, but we focus our experiments mainly on \textit{Graph Attention Network} (GAT) \cite{conf3}, which shows superior performance w.r.t. the other GNNs for the task of artist similarity. GAT leverages the multi-headed self-attention mechanism, which helps to measure how much a neighbor attends to another node. This is a huge advantage when dealing with the heterophily of the graph at hand. 

\begin{figure}[t]
    \centering
    \includegraphics[width=0.35\textwidth,keepaspectratio]{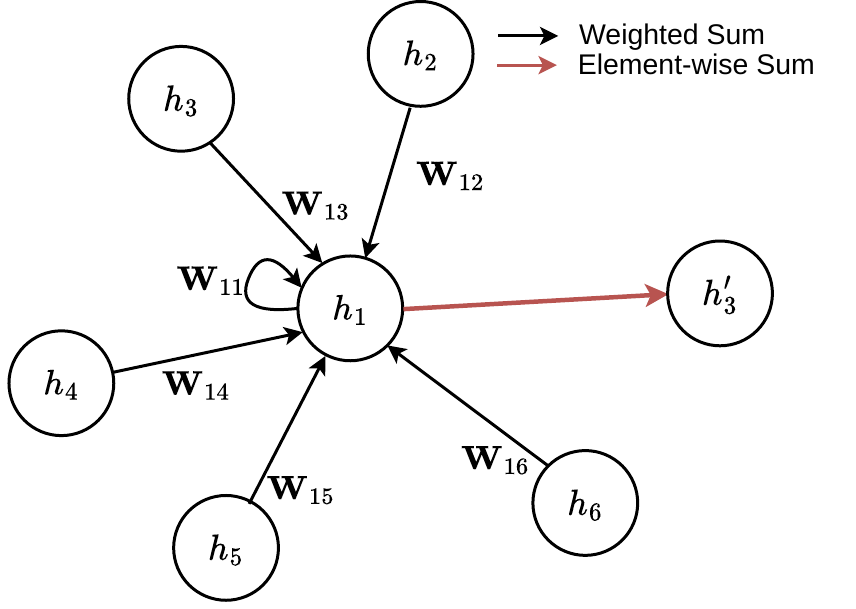}
    \caption{GAT attention mechanism, where $\mathbf{W}_{ij} = \alpha_{ij}\textbf{W}$.}
    \label{fig:scheme}
\end{figure}

According to \cite{conf3}, the attention score for a generic pair of artists $i$ and $j$ at the $l$-th layer is computed as:
\begin{equation}
    \label{eq:attentionscore}
    \alpha_{ij}^{l} = \frac{\exp(f([\textbf{W}^lh_i^l \| \textbf{W}^lh_j^l]))}{\sum_{k \in \mathcal{N}_i} \exp(f([\textbf{W}^lh_i^l \| \textbf{W}^lh_k^l])) }.
\end{equation}
where $l \in \{0, ..., L\}$, $h_i^l \in \mathcal{R}^{F_l}$ represents the hidden representation of node $i$, $\textbf{W}^l \in \mathcal{R}^{F_{l+1} \times F_l}$ denotes a tensor of trainable parameters, and $f: \mathcal{R}^{2F_{l+1}}\rightarrow\mathcal{R}$ is a simple feedforward neural network which performs the attention operation between $i$ and $j$. The attention score is computed for each pair of artists that belong to the same neighborhood. In the undirected graph case, we have always that $i \in \mathcal{N}_j$ and $j \in \mathcal{N}_i$. 
These scores are then used to retrieve the new node representations. This process is underlined by:
\begin{equation}
\label{eq:message_pass}
    h_i^{l+1} = \sigma(\sum_{k \in \mathcal{N}_i} \alpha_{ik}^{l} \textbf{W}^lh_k^l).
\end{equation}
\noindent where $h_i^{l+1} \in \mathcal{R}^{F_{l+1}}$ are the higher level features for artist $i$, and $\sigma(\cdot)$ is a nonlinearity. As previously stated, each new node representation results in a weighted combination of its neighbors. A graphical scheme of this methodology is depicted in Fig.~\ref{fig:scheme}.
The hidden representations $h_i^{l}$ for each artist can be grouped into a matrix $\mathcal{H}^l \in \mathcal{R}^{n \times F_{l}}$. We consider $\mathcal{H}^0 \equiv \mathcal{X}$ and $\mathcal{H}^L \equiv \mathcal{Z}$.

\subsection{Loss Functions}
To optimize the final embedding space such that similar artists are close to each other, we exploit the triplet loss function \cite{triplet1}:
 \begin{equation}
 \label{eq2}
    \mathcal{L}_{\text{tri}}\left(z_a,z_p,z_n\right) = \left[d\left(z_a,z_p\right)-d\left(z_a,z_n\right)+\Delta\right]^+.
\end{equation}

\noindent In \eqref{eq2}, ${z_a, z_p, z_n} \in \mathcal{Z}$ represent the embeddings for the anchor, the positive, and the negative sample, respectively, $d(\cdot, \cdot)$ is the Euclidean norm, $\Delta$ is a margin value, and $[\cdot]^+$ denotes the rectified linear unit (ReLU) function.

The loss function minimizes the distance between the anchor and its positive while maximizing the distance from its negative. In this scenario, we defined positive samples as nodes in the same neighborhood and negative ones otherwise. We used the same approach illustrated in \cite{triplet3} to compute the triplets based on distance-weighted sampling. During the training phase, we adopted a minibatch setting as described in \cite{inductive}, while the triplets were chosen within the respective minibatch. This contrastive optimization approach is then coupled with a cross-entropy loss $\mathcal{L}_{\text{CE}}(\cdot)$ in the case of label availabilities. The cost function, when including the supervised objective, is formalized as:
\begin{equation}
\label{eq:objective}
    \mathcal{L}(\mathcal{Z}, \mathcal{Y}) = \sum_{{z_a, z_p, z_n} \in \mathcal{Z}}\mathcal{L}_{\text{tri}}(z_a,z_p,z_n)  + \mathcal{L}_{\text{CE}}(\hat{\mathcal{Y}}, \mathcal{Y}).
\end{equation}



\subsection{Model Architecture}
\label{sec:moddesc}
The proposed GATSY framework is completely suited to the problem of music artist similarity. We derived an architecture based on two main blocks. The first one consists of a set of \textit{fully connected} (FC) layers followed by a second block of graph convolution (GC) operations. This guarantees a limited model complexity while avoiding any over-smoothing. In particular, we use a set of 3 FC layers in the first block and 2 GC layers in the second one. This configuration involves, at each transformation $256$ features, and it projects the $m$ input features to a $m’$ vector space (where $m$ and $m’$ are $2613$ and $256$ respectively). We also use it as the backbone architecture for the supervised learning task, including the artist-genre classification, where we added on top of GATSY a set of 2 FC layers to predict the music genre. We use the exponential linear unit (ELU) \cite{elu} after each layer except the output one, similarly to \cite{artistS}.

\section{Experimental Setup}
\label{sec:experiments}
In this Section, we provide the information concerning our experimental setup. Our objective was to compare with \cite{artistS}, but their code is proprietary, and at the time of the experiments, the dataset was not freely available. We retrieved most of the dataset and kept a similar evaluation setting. In fact, our evaluation dataset is Olga, which is one of the most popular datasets for music similarity \cite{artistS}.,here we describe the modifications it underwent, and also the characteristics that we introduced. Later in this Section, we also report the models we included in our benchmark to compare against GATSY, which is the same as denoted in \cite{artistS}. 

Recently, the full dataset was made easily accessible, and we demonstrate that our experiments on this complete dataset lead to the same conclusions as those drawn from the reduced version.

\subsection{Original Dataset}
We perform the model evaluation over Olga \cite{artistS}, 
which is one of the most popular datasets for music similarity \cite{artistS}. Although its size is limited, it is considered one of the largest available datasets for this task.
In this work, most of the experiments are based on the available samples of Olga. In fact, it was not possible to retrieve the whole dataset, according to the instructions provided in \cite{artistS}. For this reason, we used a subgraph, whose description can be found in Table \ref{OlgaD}. 
Besides the artist relations $\mathcal{A}$, the Olga dataset shows low-level audio features or input features, denoted as $\mathcal{X}$, representing track-level statistics about loudness, dynamics, and spectral shape of the signal. Moreover, these features include more abstract descriptors of rhythm and tonal information, such as the bpm and the average pitch class profile.
The dataset is in the form of an undirected graph and we split it into training/validation/test sets in the percentages of 80/10/10, respectively.
\begin{table}[t]
    \centering
    \caption{Statistic description of the original Olga, the available Olga and for our new augmented and labeled version of Olga.} 
    \begin{tabular}{c c c c}
    \hline
    \textbf{Statistics} &\textbf{Original} &\textbf{Available} & \textbf{Labelled}\\  \hline
    {Tot. n\textdegree of nodes} & 17,673 & 11,261& 11,261 \\ \hline
    {Tot. n\textdegree of connections} & 101,029& 63,096& 62,982 \\ \hline
    {Avg. n\textdegree of connections/artist} & 11.43& 11.20 & 11.46\\  \hline 
    
    {1st quartile} &3 & 3& 4 \\ \hline
    {2nd quartile} & 7& 7& 8 \\ 
    \hline {3rd quartile}&16 & 16 & 16 \\ \hline 
    \end{tabular}
    \newline

    \label{OlgaD}
\end{table}

\subsection{New Augmented Version of the Dataset}
One of the missing aspects of this dataset, in addition to its limited size, is the absence of labels associated with the artists. In fact, a supervised approach, together with the information offered by the graph structure, would lead to a model that is more aware of the cultural aspects behind the representation of the artists. 

Driven by this need, we structured a new version of Olga containing genre labels for each artist in the graph. In the reduced Olga dataset, we attempted to retrieve the music genres through the MusicBrainz\footnote{\url{https://musicbrainz.org/}} API. 
However, we were not able to retrieve labels for all the artists, so we also based this retrieval phase using NLP strategies. 

MusicBrainz is an online library of music artists through which metadata generated by users using the service can be extracted. This information often includes music genres that the users themselves have associated with each artist. This association between these music genres and the samples in the dataset was possible via the MusicBrainz IDs already available in the original Olga dataset.

In this new version of the dataset, we explicitly list all the distinct $2842$ genres we found for the artists. However, to carry out an admissible and interpretable analysis, we wanted to consider only the $25$ most relevant and frequent ones, assuming that the excluded labels are related sub-genres. Unfortunately, it was not always possible to have a unique musical genre for each artist, so disambiguation strategies were implemented as described below.
\begin{itemize}
    
    \item{Artists with multiple music genres were identified with the music genre most voted by users.}
    
    \item{Artists with a music genre that was not among the $25$ chosen ones, were labeled via a RoBERTA-based model \cite{roberta} from the \textit{spacy-sentence-bert} library. The approach used for the association is described by the following equation:
    \begin{equation}
        \label{eq:transformer}
        \arg \max_{g \in \mathcal{Y}} {\cos\left(T\left(\hat{g}, a\right), T\left(g, a\right)\right)}
    \end{equation}

    \noindent where $\hat{g}$ is the out-of-distribution genre, and $\cos\left(\cdot, \cdot\right)$ is the cosine similarity function. $T\left(\hat{g}, a\right)$ is a pre-trained model that takes as input a prompt in the form of: 
    
    \begin{center}\textit{$\hat{g}$ is the genre played by the artist $a$}\end{center}}
    
    \item{Artists without an associated tag were identified with the most common musical genre among their neighbors in the graph.}
    
    \item{Artists related to the $275$ disconnected nodes were discarded from the dataset. The new statistics of Olga, for the supervised modality, are specified in the ``New dataset" column of Table \ref{OlgaD}.}
\end{itemize}

The new version of the dataset allows for extending the set of experiments, including not only an unsupervised regime but also a supervised one. In other words, we lead two sets of experiments: one is unsupervised and exploits only the loss function in Equation \eqref{eq2}, while the second set of experiments encompasses also a supervised objective according to Equation \eqref{eq:objective}. In both settings (supervised and unsupervised), the accuracy results have been tested with 10 different random seeds and show similar performance. Our goal is to improve the quality of our recommendation system by leveraging the new characterization of the nodes in the graph. 
\subsection{Homophily of the Labeled Olga Dataset} \label{sec
}
\label{sec:homophily}
Another key characteristic of this dataset is the concept of homophily. Homophily refers to the tendency of nodes to connect with other nodes that are similar to themselves. This concept has been widely studied in the context of node classification and the performance of GNNs on datasets with low homophily, also known as heterophily \cite{improvingassortativity_heterophily, acm_heterophily, neuralsheaf_heterophily}. Typically, homophily is measured based on the classes of the nodes. These metrics indicate low homophily (i.e., heterophily) when there are many connections between nodes of different classes.

In our case, we have a graph of artists with labels that approximate their music genres. We expect the graph to exhibit heterophilic behavior because the connections between artists are not necessarily driven by music genre similarity. Following the approach in \cite{edge_homophily}, we measure homophily using \textit{edge homophily}, which represents the proportion of edges that connect nodes of the same class. It is defined as:

\begin{equation} \label{eq
} \xi_{edge} = \frac{|(i, j) \in \mathcal{E} : g_i = g_j|}{|\mathcal{E}|}, \quad g_i \in \mathcal{Y}, \ g_j \in \mathcal{Y} \end{equation}

In the labeled Olga dataset, the edge homophily value is $\xi_{edge} = 0.4757$, indicating that the majority of edges connect nodes of different classes. Although more recent and accurate measures of homophily have been proposed in the literature, they are beyond the scope of this paper. For a deeper exploration of these measures, we refer interested readers to \cite{criticallook}.
\subsection{State-of-the-Art Models for Comparisons}
In order to prove the effectiveness of the proposed GATSY network, we compare its performance with state-of-the-art models. In particular, we take into account GraphSAGE with three GC layers (for comparison purposes) from \cite{artistS} and different configurations based on GAT. We also conduct experiments with a simple FC network to evaluate the GC layer contribution. 
%

Conversely, from \cite{artistS}, we also included batch normalization layers \cite{batch} after each transformation, except for the final one. This choice has been favoured by the advantages that it entails, mostly in terms of accuracy. The details on the hyperparameters are available in our codebase.
%
%


\subsection{Evaluation metrics}
\label{accuracy}
The embedding evaluation is carried out with the normalized discounted cumulative gain (nDCG) metric \cite{nDCG, artistS}. The procedure is explained as follows.
We first compute the embedding of artists using all the connections between the nodes in the training and evaluation set. Then, for each sample in the latter, all the mutual distances are computed.  
The distances are ranked, and for each instance, we consider its $K$ nearest neighbors. If its neighbors are also connected to the instance, then the prediction is recognized as a true positive.


For our experiments, we set $K=200$. Further details on the evaluation setting are provided in \cite{artistS}.
For the prediction of the classes for each artist, we employ the f1-score. This, of course, only holds for the supervised learning experiments.

\begin{figure}[t]
    \centering
    \includegraphics[width=0.38\textwidth,keepaspectratio]{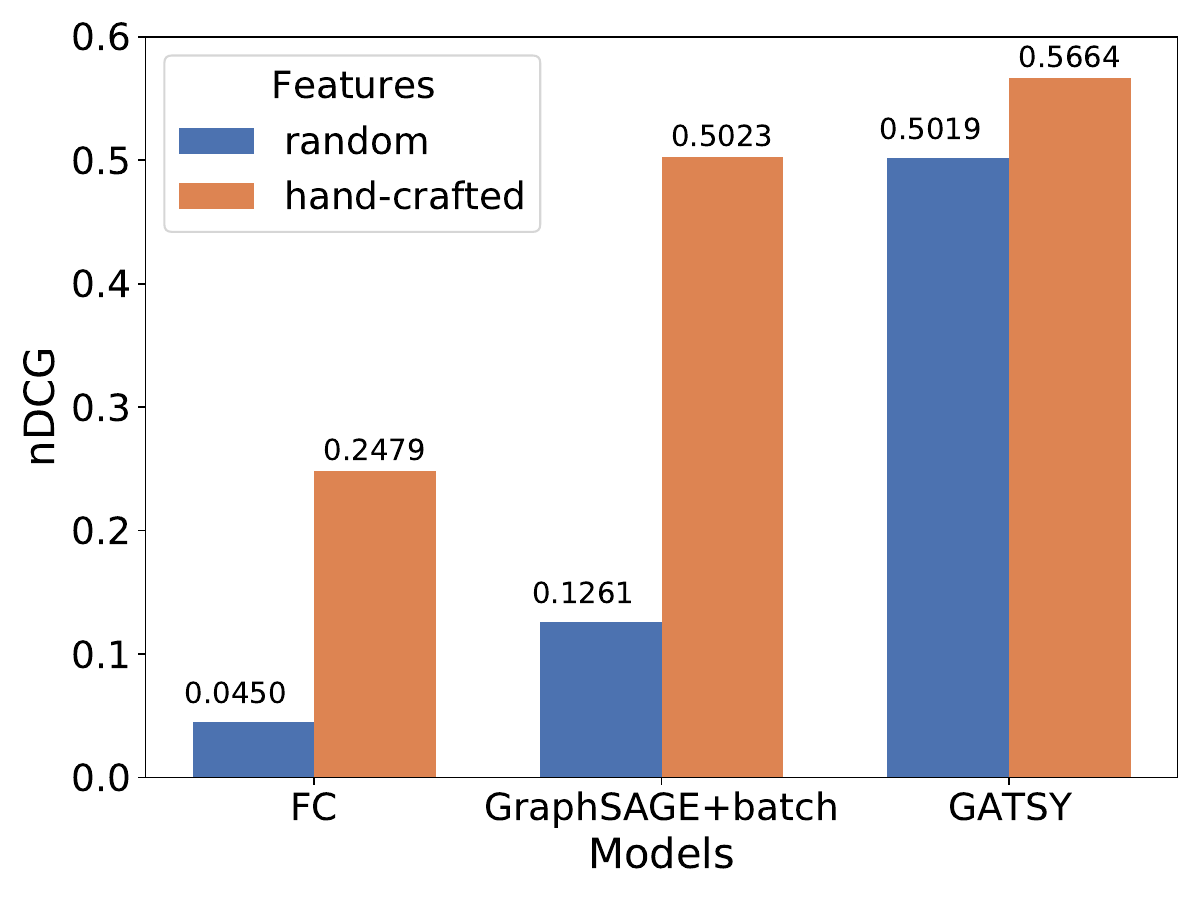}
    \caption{Comparison between the low-level and random features experiments.}
    \label{fig:ablation}
\end{figure}

\section{Results}
\label{sec:results}
In this section, we provide the results of our main experiments, emphasizing the performance of our best settings. We analyze the contribution of the key components of our architecture through an ablation study to assess their importance. Using data visualization tools, we demonstrate how the network organizes the embedding space into clusters of artists. Finally, we evaluate the effectiveness of this approach as a recommendation system, and we propose a method to introduce fictitious artists into the dataset, showing how they can be easily related to existing artists.

Our research questions guiding these experiments are as follows:

\begin{itemize}
    \item \textbf{RQ1:} How does GATSY perform in terms of artist similarity compared to previous baselines and other GNNs?
    \item \textbf{RQ2:} What is the cost of our approach in terms of model space complexity?
    \item \textbf{RQ3:} What are the effects of using both structure and input features on the performance of our model?
    \item \textbf{RQ4:} Are class labels helpful to improve performance on unsupervised metrics?
    \item \textbf{RQ5:} How can the similarity computation be effectively applied in real-world scenarios, including the introduction of fictitious artists?
\end{itemize}

\subsection{Artist Similarity}

According to the procedures described in Section \ref{sec:experiments}, we present in Table \ref{tab:results} the main findings of our method compared to the approach proposed in \cite{artistS}. Specifically, the 	\textbf{Reduced Dataset} box reports the results obtained on the dataset we have focused on throughout our research.

With the recent release of the full version of Olga, as specified in Table \ref{OlgaD}, we have extended our experiments to the complete dataset. These results are presented in the 	\textbf{Full Dataset} box. We conducted these experiments using four different GNN architectures—GCN, GIN, GraphSAGE, and GAT—implementing them following the algorithm described in \cite{artistS}. We performed extensive hyperparameter tuning for all models and trained them until convergence.

In the 	\textbf{Reduced Dataset} box, our GAT-based model, which we name GATSY, exhibits superior performance compared to all competitors. While GraphSAGE benefits from batch normalization to enhance its performance significantly, GATSY still outperforms it by a notable margin. Furthermore, GATSY demonstrates advantages in model complexity and efficiency, requiring fewer layers and parameters than GraphSAGE.

The 	\textbf{Full Dataset} box provides additional insights into our research questions by presenting results on the original Olga dataset, which includes all connections and artists from the full graph. Here, we repeated the unsupervised task experiments, comparing GATSY not only with GraphSAGE but also with other GNNs, such as GCN \cite{conf1} and GIN \cite{GIN}. Notably, we observe that GraphSAGE underperforms compared to the results reported in \cite{artistS}. We attribute this discrepancy to the differences in implementation, as we could not access the original proprietary code and had to reimplement the method from scratch. Despite this limitation, GATSY continues to outperform all baselines across both datasets, reinforcing the effectiveness of graph attention for the artist similarity task. We highlight this by naming our GAT-based model GATSY. Additionally, both GAT and GCN stand out as more self-contained models in terms of parameter efficiency.

Table \ref{tab:results} and our analysis contribute to answering 	\textbf{RQ1} and \textbf{RQ2}.
\begin{figure}
    \centering
    \includegraphics[width=0.98\linewidth]{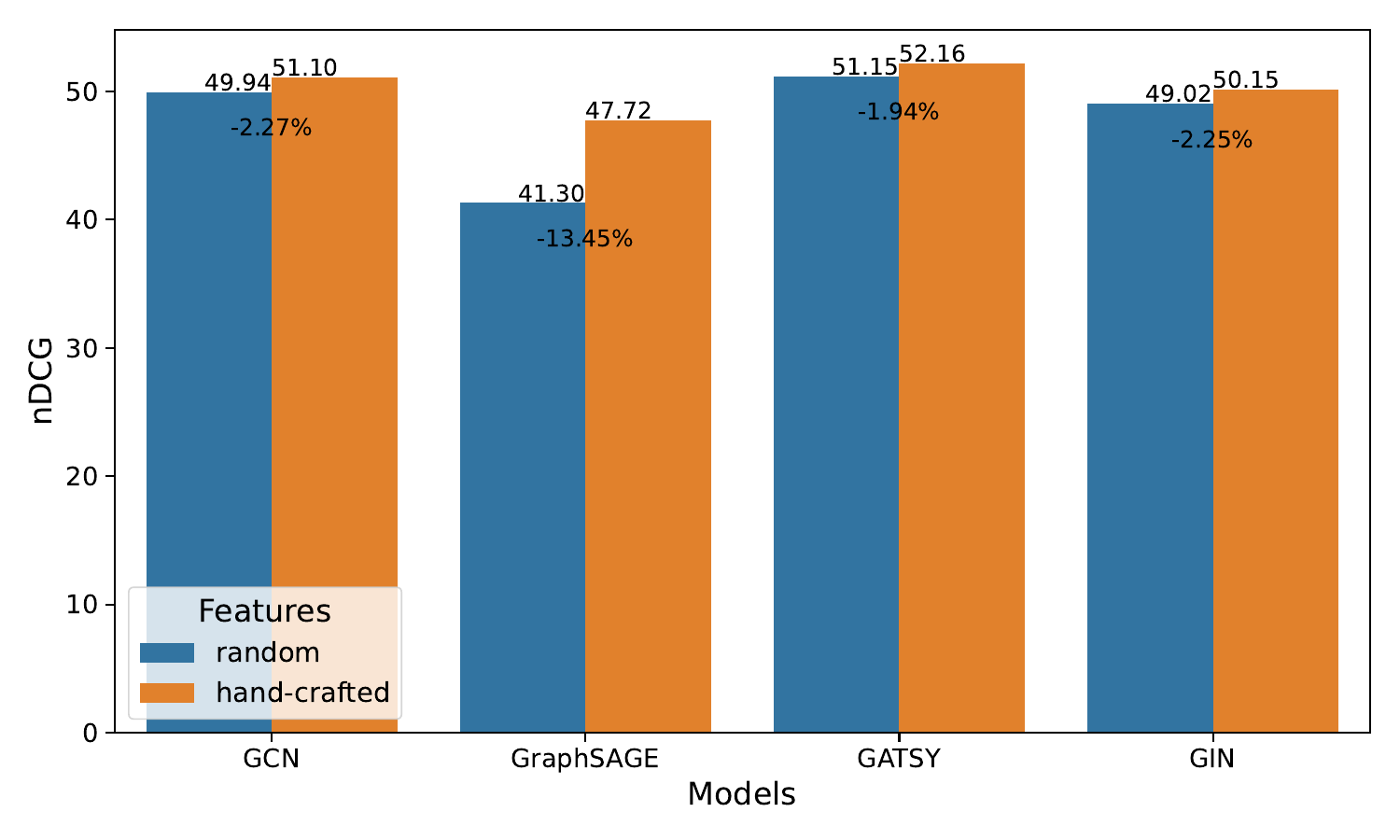}
    \caption{GNN model comparisons on the full Olga datasets. GATSY is the most resilient to random features among the other baselines.}
    \label{fig:results_full_olga}
\end{figure}
\begin{table}[t]
    \centering
    \caption{Results gained from the main experiments over 10 trials. We report the mean value with the related standard deviation. Accuracy refers to nDCG.}
    \begin{tabular}{c c c}
    \hline
        \textbf{Model} & \textbf{Accuracy} &  \textbf{N\textdegree of parameters} \\ \hline
        \multicolumn{3}{c}{\textbf{Reduced Dataset}} \\ \hline
        FC & 0.2479 ± 0.004 &  800,768 \\ 
        GraphSAGE  & 0.4180 ± 0.0052 & 1,758,052 \\  
        + Batch Norm. & 0.5023 ± 0.0048 & 1,759,588 \\  
        GATSY      &  \textbf{0.5664 ± 0.0068} & 936,448 \\ \hline
        \multicolumn{3}{c}{\textbf{Full Dataset}} \\ \hline
        GCN & 0.5110 ± 0.0038 & 932,352 \\  
        GraphSAGE & 0.4772 ± 0.0098 & 8,720,018 \\  
        GATSY & \textbf{0.5216 ± 0.0055} & 933,888 \\  
        GIN & 0.5015 ± 0.0046 & 998,144 \\ \hline
    \end{tabular}
    \label{tab:results}
\end{table}

\subsection{Ablation Study}
\label{ablation}
To understand the key components of our method that contribute most to its overall performance, we refer to the ablation study presented in Figure \ref{fig:ablation} (reduced dataset) and Figure \ref{fig:results_full_olga} (full dataset). They both present, across all the baselines, how performance drops depending on the type of input features used—whether random or hand-crafted. In Figure \ref{fig:ablation}, in the FC and GraphSAGE models, we observe a similar phenomenon as reported in \cite{artistS}, despite the changes in the graph structure (see Table \ref{OlgaD}), indicating that GraphSAGE is highly dependent on input features. The performance of the FC model, on the other hand, reinforces the idea that incorporating graph structure would enhance model performance.
In the case of GATSY, Fig.~\ref{fig:ablation} shows that the loss in performance is relatively low compared to the baselines. Specifically, the performance loss is 81.8\% for FC, 74.89\% for GraphSAGE, and only 11.38\% for GATSY. Once, we had access to the full Olga dataset, we also repeated this very same ablation study, considering also the other GNN models. The results are reported in Figure   \ref{fig:results_full_olga}. Also within the full dataset, we can assess that GATSY is the solution most resilient to random features, and GraphSAGE behaves as the one struggling the most. Also GCN and GIN seem resilient to input features as well, but their performance is slightly lower than GATSY. 

These results demonstrate that GATSY is less affected by the absence of hand-crafted features. This analysis addresses \textbf{RQ3}, confirming that GATSY effectively utilizes structural and input features, making it more robust than other models when hand-crafted features are missing.

\begin{figure*}[t!]
    \centering
    \includegraphics[width=0.86\textwidth,keepaspectratio]{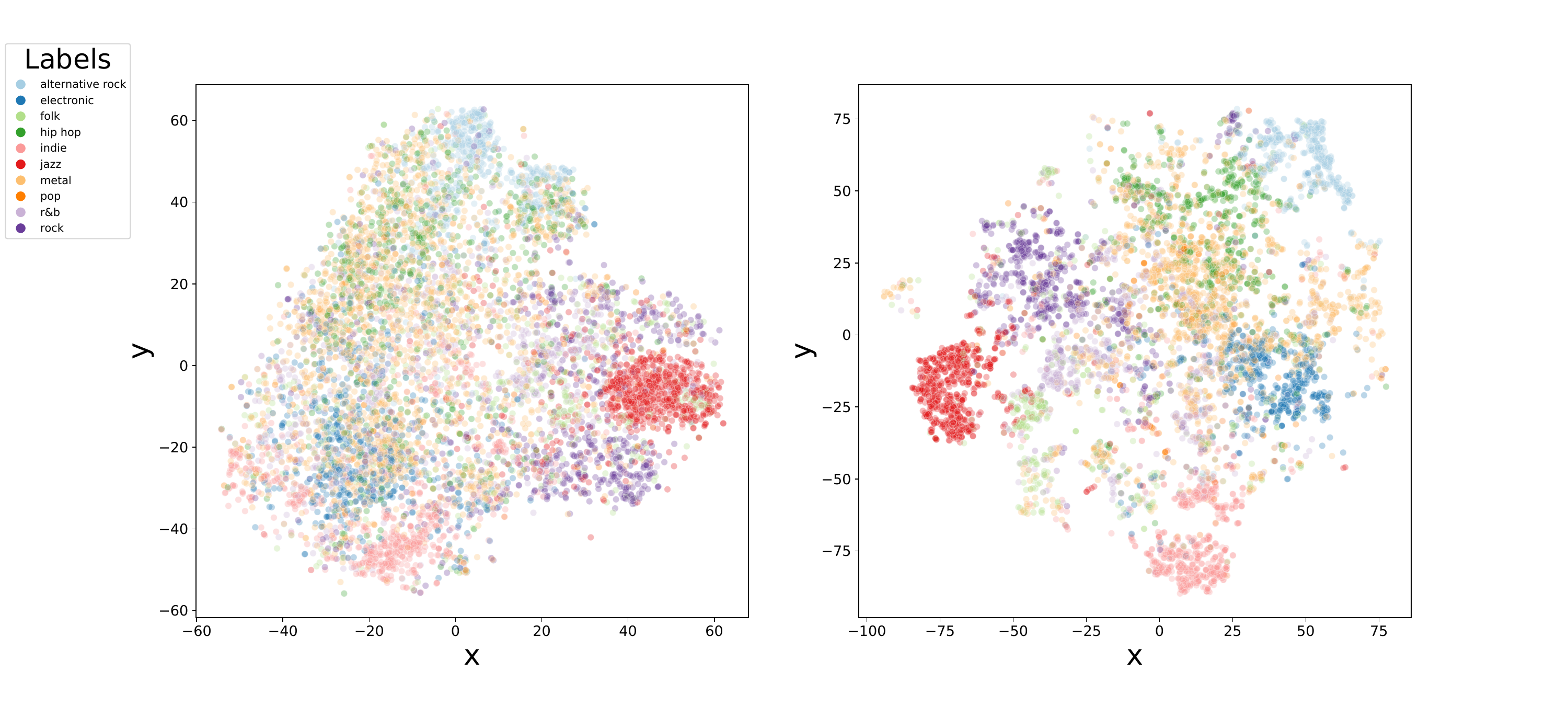}
    \caption{Visualization of the low-level features after dimensionality reduction (LEFT), and of the GATSY embedding after training on labeled Olga and dimensionality reduction (RIGHT). Here are shown the embeddings labelled among only the top 10 genres for visualization purposes. }
    \label{fig:labeled_embs}
\end{figure*}
\begin{table}[t]
    \centering
    \caption{Results gained from the experiment with the reduced dataset. We report the mean value with the related standard deviation and the F1-score for artist genre classification. Accuracy refers to nDCG.}
    \begin{tabular}{c c c}
    \hline
        \textbf{Model} & \textbf{Accuracy} &  \textbf{F1-Score} \\ \hline
        GATSY      &  0.5682
 ± 0.0051
 & - \\ \hline
 GATSY+FC & 0.5583 ± 0.0070 & 0.5445 ± 0.0096\\ \hline 
         
    \end{tabular}

    \label{tab:results2}
\end{table}
\subsection{Supervised Learning Experiments}
In this Section, we describe the performance achieved in the supervised experiments. Here, we rely on the labelled dataset, where the number of nodes is further reduced (see Table \ref{OlgaD}). We structure new splits and provide a new baseline for the unsupervised case (see GATSY in Table \ref{tab:results2}). This way, we can show the performance variation of the model when the objective becomes the one in Equation \eqref{eq:objective}.
In Table \ref{tab:results2} we show the performance of the genre-aware model. We confirm that the $\text{nDCG}_{200}$ is more expressive as an unsupervised metric. Indeed, the unsupervised approach is always outperforming the supervised experiments in terms of accuracy. Despite this, the supervised trained GATSY becomes capable of predicting the artist labels, which means that it is more genre-aware than the unsupervised trained GATSY. Furthermore, we expect that achieving a high F1-score is quite challenging, as associating an artist with a unique genre is rather unfeasible. This is also the reason why the supervised loss struggles to help the unsupervised one. For instance, in the Olga graph, the neighborhood is identified by a unique class, as corroborated by our discussion on the homophily level of Olga (see Section \ref{sec:homophily}). In this sense, the supervised and unsupervised objectives may be colliding whenever two neighbors of different classes attract themselves to minimize $ \mathcal{L}_{\text{tri}}$, and also repulse to minimize $\mathcal{L}_{\text{CE}}$. These conclusions allow us to answer negatively to \textbf{RQ4}. 

In Fig.~\ref{fig:labeled_embs}, we show the low-level features associated with the music genre, and the clusters we obtain when GATSY is trained both on $\mathcal{L}_{\text{tri}}$ and $\mathcal{L}_{\text{CE}}$. Here, the embeddings organize themselves in the space driven by the supervision signal associated with the music genre. The clusters are identified by the ground-truth values in $\mathcal{Y}$.
\subsection{Artist Recommendation}
\label{recom}
To answer \textbf{RQ5}, we now describe how our method and, in general, the artist similarity task can work as a recommender system. To enhance heterogenous recommendations in terms of music genres, in this Section we show some of the results we had from the unsupervised experiments. These results also provide a qualitative perspective on how these systems work in practice.\\
\textbf{Recommendation retrieval.}
In the following, we describe the procedure to get the $k$-nearest-neighbors from a query artist:
\begin{enumerate}
	\item{Compute the embedding of artists with the complete adjacency matrix and their respective feature vectors;}
	\item{Select the latent vector for the query artist;}
	\item{Compute its $k$-nearest-neighbors in the embedding and see what artists represent.}
\end{enumerate}
\noindent In Table \ref{tab:gat} and Table \ref{tab:nograph}, we report some of the recommendations obtained through this procedure. These are obtained using the reduced dataset. \\
\begin{table}[t]
    \caption{These are the 5-nearest-neighbors retrieved from the embedding of the GATSY.}
    \centering{
    \begin{tabular}{|c|c|c|}\hline
         \multicolumn{3}{|c|}{\textbf{GATSY}}  \\ \hline
        {\textbf{Snoop Dogg}}& {\textbf{Nancy Sinatra}}&
        {\textbf{Rod Stewart}}\\ \hline
        
         Warren G  & Cilla Black  & Fleetwood Mac \\ \hline 
         2Pac  & Harpers Bizarre  & 
         Dire Straits\\ \hline
         Tha Dogg Pound &  Sonny \& Cher &  Paul McCartney \\ \hline
         Spice 1  & Helen Shapiro  & Ringo Starr \\ \hline 
         Luniz  & Sandie Shaw  & Bruce Hornsby  \\ \hline
    \end{tabular}}

    \label{tab:gat}
\end{table}
\begin{table}[t]
    \caption{These are the 5-nearest-neighbors retrieved from the embedding when the network does not use graph convolutional layers.}
    \centering{
    \begin{tabular}{|c|c|c|}\hline
          \multicolumn{3}{|c|}{\textbf{No Graph Layers}}  \\ \hline
         {\textbf{Snoop Dogg}}& {\textbf{Nancy Sinatra}}&
        {\textbf{Rod Stewart}}\\ \hline
        
         Afu-Ra  & Dan Penn  & John Entwistle \\ \hline 
         Erik Sermon  & Cass Elliot  & Ty Segall \\ \hline
         Camp Lo &  Paper Lace & Blitzen Trapper \\ \hline
         Wu-Tang Clan& Rosie Flores & Charlie Sexton\\ \hline 
         Swollen Members  & Jerry Butler  & Ugly Kid Joe  \\ \hline
    \end{tabular}}
    \label{tab:nograph}
\end{table}
\textbf{Recommendation for fictitious artists.} One of the novelties of our work relies on the possibility of adding fictitious artists like in \cite{spinelli}, to create a bridge between two possibly distant genres and still obtain plausible suggestions. We trust this procedure thanks to the robustness of GATSY to the input features.

Let us assume that we want to augment the dataset with an artist who is not real. We need to provide it with a feature vector $x_{fic}$ and add a row to the adjacency matrix. The new row is a binary vector {$a_{i}$} defined as follows:
\begin{equation}
\begin{cases}
a_{i,j} = 1          $ \textit{  if j is the index for an artist in }$ \mathcal{S} \\ 
a_{i,j} = 0           $ \textit{  otherwise}$
\end{cases}
\end{equation}

\noindent where $\mathcal{S}$ is the set of artists similar to the non-existing artist.
We can also define its node features as the average of the features related to the artists in $\mathcal{S}$.
Defining $\mathcal{A}_{a}$, $\mathcal{X}_{a}$ as the augmented adjacency, we obtain the embedding of the fictitious artist as

\begin{equation}
\mathcal{Z}_a = \text{GNN}(\mathcal{X}_{a}, \mathcal{A}_{a}).
\end{equation}

To illustrate the performances with the non-existing artists, we have conducted experiments with different mixings in $\mathcal{S}$. In particular, they were considered combinations among the same or similar music genres, referred to as \textit{Likely mixings} (Table \ref{tab:likely}), and very different ones \textit{Unlikely mixings} (Table \ref{tab:unlikely}). As expected, the likely mixings keep a consistency, in terms of music genres, all over the recommendations, which is not the case for the unlikely ones. Interestingly in the latter scenario, the recommendations converge on ``Dance/Jazz" artists. The last artist, Arthur Baker, is also a famous hip-hop/rap producer, which corroborates the validity of our recommendation.
\begin{table}[!t]
    \caption{Lixely mixing example. 'Dire Floyd' artists in $\mathcal{S}$: [ 'Pink Floyd', 'Dire Straits', 'Jethro Tull']}
    \centering{
    \begin{tabular}{|c|c|}
    \hline
        \textbf{Name}& \textbf{Genre}   \\ \hline
     Roger Waters  & Prog Rock \\ \hline 
 Greg Lake &  Prog Rock\\ \hline
 Procol Harum & Prog Rock \\ \hline
  Adrian Belew& Prog Rock \\ \hline 
Golden Earring& Rock\\ \hline 
    \end{tabular}}

      \label{tab:likely}
\end{table}

\begin{table}[!t]
\caption{Unlikely mixing example. 'Classic/Rap' artists in $\mathcal{S}$: ['Claudio Abbado', 'Edvard Grieg', 'Riccardo Muti', 'Kurt Masur', 'Herbert von Karajan', 'Yo-Yo Ma', '2Pac', 'Snoop Dogg', 'Ice Cube', 'Eminem', 'JAY‐Z']}
    \centering{
    \begin{tabular}{|c|c|}
    \hline
        \textbf{Name} &   \textbf{Genre}   \\ \hline
     André Previn &   Jazz \\ \hline 
Jean‐Michel Jarre &  Electronic\\ \hline
 Nile Rodgers & Dance \\ \hline
 Quincy Jones & Jazz \\ \hline 
Arthur Baker &Dance\\ \hline
\end{tabular}}

\label{tab:unlikely}
\end{table}

%
%
%
%
\section{CONCLUSION}
\label{sec:conclusion}
In this work, we addressed the task of music artist similarity by proposing GATSY, a graph neural network framework that is robust to the absence of input features, and more expressive and efficient than previous approaches. We developed an augmented version of the Olga dataset that now includes labels for the artists, allowing us to enhance the unsupervised dataset and implement a genre-aware model. However, we provide evidence that using genre-based supervision does not improve the artist similarity performance, as two artists can be similar even if they come from different genres, and also because it is hard to identify a unique genre for a music artist. We also showed in Section \ref{recom} an artist recommendation pipeline that fits more within real-world scenarios. We introduced the possibility of creating ``bridge" nodes to connect unliked artists. By performing this data augmentation via node injection, it is possible to connect two or more previously unrelated artists and generate recommendations based on the new topology.
Many challenges still need to be solved by the research community. In future works, it would be interesting to assess the capabilities of GATSY in long-tail recommendation and evaluate whether using random features when hand-crafted features are not available could be part of the solution. Moreover, other challenges, like the absence of benchmark datasets, should be of interest to the community. Even the development of a more effective and scalable model that can satisfy the demand coming from modern music streaming services.
%

\bibliographystyle{IEEEtran} 
\bibliography{main}

\end{document}